\documentclass[twocolumn,showpacs,preprintnumbers,amsmath,amssymb,superscriptaddress,unsortedaddress]{revtex4}

\usepackage{graphicx}
\usepackage{dcolumn}
\usepackage{bm}

\begin{document}

\preprint{APS/123-QED}

\title{Neutron skin of $^{208}$Pb from Coherent Pion Photoproduction}

\author{C.M.~Tarbert}
\author{D.P.~Watts}%
 \email{dwatts1@ph.ed.ac.uk}
\affiliation{SUPA, School of Physics, University of Edinburgh, Edinburgh, United Kingdom}

\author{D.I.~Glazier}
\affiliation{SUPA, School of Physics, University of Edinburgh, Edinburgh, United Kingdom}

\author{P.~Aguar}
\affiliation{Institut f\"ur Kernphysik, University of Mainz, Germany}

\author{J.~Ahrens}
\affiliation{Institut f\"ur Kernphysik, University of Mainz, Germany}

\author{J.R.M.~Annand} 
\affiliation{SUPA, Department of Physics and Astronomy, University of Glasgow, Glasgow, United Kingdom}

\author{H.J.~Arends}
\affiliation{Institut f\"ur Kernphysik, University of Mainz, Germany}

\author{R.~Beck}
\affiliation{Institut f\"ur Kernphysik, University of Mainz, Germany}
\affiliation{Helmholtz-Institut f\"ur Strahlen- und Kernphysik, University Bonn, Germany}

\author{V. ~Bekrenev}
\affiliation{Petersburg Nuclear Physics Institute, Gatchina, Russia}

\author{B.~Boillat} 
\affiliation{Institut f\"ur Physik, University of Basel, Basel, Ch}

\author{A.~Braghieri} 
\affiliation{INFN Sezione di Pavia, Pavia, Italy}

\author{D.~Branford}
\affiliation{SUPA, School of Physics, University of Edinburgh, Edinburgh, UK}

\author{W.J.~Briscoe}
\affiliation{Center for Nuclear Studies, The George Washington University, Washington, DC, USA}

\author{J.~Brudvik}
\affiliation{University of California at Los Angeles, Los Angeles, CA, USA}

\author{S.~Cherepnya}
\affiliation{ Lebedev Physical Institute, Moscow, Russia}

\author{R.~Codling} 
\affiliation{SUPA, Department of Physics and Astronomy, University of Glasgow, Glasgow, UK}

\author{E.J.~Downie} 
\affiliation{SUPA, Department of Physics and Astronomy, University of Glasgow, Glasgow, UK}

\author{K.~Foehl}
\affiliation{SUPA, School of Physics, University of Edinburgh, Edinburgh, UK}

\author{P.~Grabmayr}
\affiliation{Physikalisches Institut Universit\"at T\"ubingen, T\"ubingen, Germany}

\author{R.~Gregor} 
\affiliation{ II. Physikalisches Institut,  University of Giessen, Germany}

\author{E.~Heid}
\affiliation{Institut f\"ur Kernphysik, University of Mainz, Germany}

\author{D.~Hornidge}
\affiliation{Mount Allison University, Sackville, NB, Canada} 

\author{O.~Jahn}
\affiliation{Institut f\"ur Kernphysik, University of Mainz, Germany}

\author{V.L.~Kashevarov} 
\affiliation{ Lebedev Physical Institute, Moscow, Russia}

\author{A.~Knezevic}
\affiliation{Rudjer Boskovic Institute, Zagreb, Croatia}

\author{R.~Kondratiev}  
\affiliation{Institute for Nuclear Research, Moscow, Russia}

\author{M.~Korolija}
\affiliation{Rudjer Boskovic Institute, Zagreb, Croatia}

\author{M.~Kotulla}
\affiliation{Institut f\"ur Physik, University of Basel, Basel, Ch}

\author{D.~Krambrich}
\affiliation{Institut f\"ur Kernphysik, University of Mainz, Germany}
\affiliation{Helmholtz-Institut f\"ur Strahlen- und Kernphysik, University Bonn, Germany}

\author{B.~Krusche}
\affiliation{Institut f\"ur Physik, University of Basel, Basel, Ch}

\author{M.~Lang}
\affiliation{Institut f\"ur Kernphysik, University of Mainz, Germany}
\affiliation{Helmholtz-Institut f\"ur Strahlen- und Kernphysik, University Bonn, Germany}

\author{V.~Lisin}
\affiliation{Institute for Nuclear Research, Moscow, Russia}

\author{K.~Livingston} 
\affiliation{SUPA, Department of Physics and Astronomy, University of Glasgow, Glasgow, UK}

\author{S.~Lugert}
\affiliation{ II. Physikalisches Institut,  University of Giessen, Germany}

\author{I.J.D.~MacGregor} 
\affiliation{SUPA, Department of Physics and Astronomy, University of Glasgow, Glasgow, UK}

\author{D.M.~Manley}
\affiliation{Kent State University, Kent, OH, USA}

\author{M.~Martinez}
\affiliation{Institut f\"ur Kernphysik, University of Mainz, Germany}

\author{J.C.~McGeorge} 
\affiliation{SUPA, Department of Physics and Astronomy, University of Glasgow, Glasgow, UK}

\author{D.~Mekterovic}
\affiliation{Rudjer Boskovic Institute, Zagreb, Croatia}

\author{V.~Metag}
\affiliation{ II. Physikalisches Institut,  University of Giessen, Germany}

\author{B.M.K.~Nefkens} 
\affiliation{University of California at Los Angeles, Los Angeles, CA, USA}

\author{A.~Nikolaev}
\affiliation{Institut f\"ur Kernphysik, University of Mainz, Germany}
\affiliation{Helmholtz-Institut f\"ur Strahlen- und Kernphysik, University Bonn, Germany}

\author{R.~Novotny}
\affiliation{ II. Physikalisches Institut,  University of Giessen, Germany}

\author{R.O.~Owens}
\affiliation{SUPA, Department of Physics and Astronomy, University of Glasgow, Glasgow, UK}

\author{P.~Pedroni}
\affiliation{INFN Sezione di Pavia, Pavia, Italy}

\author{A.~Polonski}
\affiliation{Institute for Nuclear Research, Moscow, Russia}

\author{S.N.~Prakhov}
\affiliation{University of California at Los Angeles, Los Angeles, CA, USA}

\author{J.W.~Price}
\affiliation{University of California at Los Angeles, Los Angeles, CA, USA}

\author{G.~Rosner}
\affiliation{SUPA, Department of Physics and Astronomy, University of Glasgow, Glasgow, UK}

\author{M.~Rost}
\affiliation{Institut f\"ur Kernphysik, University of Mainz, Germany}

\author{T.~Rostomyan}
\affiliation{INFN Sezione di Pavia, Pavia, Italy}

\author{S.~Schadmand}
\affiliation{ II. Physikalisches Institut,  University of Giessen, Germany}

\author{S.~Schumann}
\affiliation{Institut f\"ur Kernphysik, University of Mainz, Germany}
\affiliation{Helmholtz-Institut f\"ur Strahlen- und Kernphysik, University Bonn, Germany}

\author{D.~Sober}
\affiliation{The Catholic University of America, Washington, DC, USA}

\author{A.~Starostin}
\affiliation{University of California at Los Angeles, Los Angeles, CA, USA}

\author{I.~Supek}
\affiliation{Rudjer Boskovic Institute, Zagreb, Croatia}

\author{A.~Thomas}
\affiliation{Institut f\"ur Kernphysik, University of Mainz, Germany}

\author{M.~Unverzagt}
\affiliation{Institut f\"ur Kernphysik, University of Mainz, Germany}
\affiliation{Helmholtz-Institut f\"ur Strahlen- und Kernphysik, University Bonn, Germany}

\author{Th.~Walcher}
\affiliation{Institut f\"ur Kernphysik, University of Mainz, Germany}

\author{F.~Zehr}
\affiliation{Institut f\"ur Physik, University of Basel, Basel, Ch}


\collaboration{The Crystal Ball at MAMI and A2 Collaboration}
%


\date{\today}

\begin{abstract}

Information on the size and shape of the neutron skin on $^{208}$Pb
is extracted from coherent pion photoproduction cross sections
measured using the Crystal Ball detector together with the Glasgow tagger at
the MAMI electron beam facility. On exploitation of an interpolated fit of a 
theoretical model to the measured cross sections the half-height radius and diffuseness of the neutron distribution are found to be $c_{n}$=6.70$\pm 0.03(stat.)$~fm and 
 $a_{n}$=0.55$\pm 0.01(stat.)$$^{+0.02}_{-0.03}(sys.)$~fm respectively, corresponding to a neutron skin thickness $\Delta r_{np}$=0.15$\pm 0.03(stat.)$$^{+0.01}_{-0.03}(sys.)$~fm. The results give the first successful extraction of a neutron skin thickness with an electromagnetic probe and indicate that the skin of $^{208}$Pb has a halo character. The measurement provides valuable new constraints on both the structure of nuclei and the equation of state for neutron rich matter.  
\end{abstract}

\maketitle


Obtaining an accurate determination of the character of the neutron distribution in nuclei has proven elusive despite decades of study. This has been a long-standing and serious shortcoming in our understanding of nuclear structure. The difference between the neutron and proton distributions is often expressed as a neutron skin thickness ($\Delta r_{np}$), defined as the difference between the root mean square (rms) radii for the neutron and proton distributions. This convention was adopted as many previous measurements had little or no sensitivity to the diffuseness of the density distributions, excluding analysis based on the more familiar half-height radius and diffuseness of a 2-parameter Fermi distribution~\cite{gmitro}. State-of-the-art nuclear theories predict $\Delta r_{np}$ values for $^{208}$Pb that range from 0.05 to 0.35fm~\cite{skinmodel}, despite all being constrained by the same nuclear data. Recently, it was pointed out that the magnitude of $\Delta r_{np}$ in heavy nuclei shows a tight, model-independent correlation with the density dependence of the symmetry energy for neutron matter~\cite{modelindep,nuclear}. The character of the neutron skin, therefore, has a wide impact and the potential to give important new information on neutron star structure and cooling mechanisms~\cite{steiner,Horowitz,Zu,Steiner2,rutel}, searches for physics beyond the standard model~\cite{nstarsand,parity}, the nature of 3-body forces in nuclei~\cite{Tsang2,Schwenk}, collective nuclear excitations~\cite{centelles,Carbone,Chen,tamii} and flows in heavy-ion collisions~\cite{Li,Tsang}.



The potential impact of obtaining an accurate determination of $\Delta r_{np}$ has led to a flurry of theoretical and experimental interest in recent years. Many studies have focused on $^{208}$Pb which has a relatively well-understood structure due to the closed proton ($Z$=82) and neutron ($N$=126) shells. A goal of a $\pm$0.05~fm accuracy in $\Delta r_{np}$ is quoted~\cite{skinmodel} as the requirement to constrain the equation of state sufficiently to remove the current major ambiguities. 
A recent review of the experimental attempts to measure $\Delta r_{np}$ in  $^{208}$Pb is given by Tsang \textit{et. al.}~\cite{Tsang2}. Recent analysis of proton~\cite{zenihiro} and pion~\cite{friedman} scattering data gave $\Delta r_{np}$=0.211$\pm$0.06~fm and $\Delta r_{np}$=0.16$\pm0.07$~fm respectively. Studies of the annihilation of antiprotons on the nuclear surface~\cite{antip1,antip2} gave $\Delta r_{np}$=0.18$\pm0.04(expt.)\pm0.05(theor.)$~fm. Isospin diffusion in heavy-ion collisions gave $\Delta r_{np}$=0.22$\pm$0.04~fm~\cite{ion_diff}. Measurements of pygmy dipole resonances and electric dipole polarizabilities of nuclei~\cite{Carbone,kras,tamii} gave $\Delta r_{np}$ ranging from 0.156~fm to 0.194~fm with quoted accuracies as small as $\pm$0.024~fm, although the model dependence is still debated~\cite{Reinhard} and an accuracy of $\pm$0.05~fm is taken in Ref.~\cite{Tsang2}.



 
Experiments with electromagnetic probes~\cite{krusche_ff,Schrack:1962,magrabi} have systematic errors different from those with the strongly interacting probes described above and have the advantage of probing the full nuclear volume. However, there have been no successful measurements of the neutron distribution. A measurement using an electroweak probe has very recently been obtained in parity-violating electron scattering on nuclei (PREX), utilising the preferential coupling of the exchanged weak boson to neutrons. A first measurement at a single momentum transfer gave $\Delta r_{np}$=0.33$\pm$0.17~fm in $^{208}$Pb~\cite{prex}. 

This Letter establishes coherent photoproduction of $\pi^{0}$ mesons from  $^{208}$Pb as an accurate probe of the nuclear shape, which has sufficient sensitivity to detect and characterise the neutron skin. In the coherent reaction the target nucleus is left in its ground state, which ensures that all the nucleons contribute  coherently to the reaction amplitude. For our data at incident photon energies 180-240 MeV, where $\Delta$ excitation is the dominant mechanism, the amplitudes for neutron and proton $\Delta$ excitation are expected to be identical~\cite{krusche_ff}. The coherent $(\gamma,\pi^{0}$) cross section, therefore, determines the nucleon density distribution in much the same way that elastic electron scattering determines the charge distribution. Compared to hadronic probes that have been used to study the neutron skin, interpretation of the $(\gamma,\pi^{0}$) reaction is advantageous as it is not complicated by initial state interactions. However, final state interactions between the  $\pi^{0}$ and the nucleus are significant; they produce both a shift in the pion emission angle and a modification of the outgoing flux, which must be accurately treated in the theoretical calculation of the  $(\gamma,\pi^{0}$) cross section if reliable nuclear shape information is to be obtained. The  $\pi^{0}$-nucleus interaction varies with energy and the validity of its treatment can therefore be assessed from the consistency of the nuclear shape parameters obtained from  $(\gamma,\pi^{0}$) angular distributions at different incident photon energies. The analysis below presents data for the $E_{\gamma}$ range 180-240~MeV. Data have also been obtained from threshold up to 180 MeV but extensions to the theoretical calculation are required before these data can be used, to allow for different photon coupling to neutrons and protons. Above E$_{\gamma}$ = 240 MeV the extracted shape parameters become unreliable,  probably as a result of the rapid increase in the $\pi^{0}$-nucleus interaction in the $\Delta$ resonance region. 

Previous  $(\gamma,\pi^{0}$) measurements for $^{208}$Pb~\cite{krusche_ff,Schrack:1962} either did not use isotopically pure targets or did not achieve the precision needed to study the neutron skin, mainly because they used  $\pi^{0}$  detection systems with limited angular coverage resulting in a small detection efficiency with too large a dependence on pion energy and angle to give definitive results. In the present experiment these problems are almost completely removed by utilising a large solid angle photon detector, the Crystal Ball (CB)~\cite{cball} in conjunction with the Glasgow photon tagger~\cite{tagger} and the MAMI electron microtron~\cite{mami}. The experimental setup is described in detail in Ref.~\cite{incoh}. The tagged photon beam had a resolution of $\sim$2~MeV full width and an intensity of $\sim$2$\times10^{5}$~$\gamma$ s$^{-1}$~MeV$^{-1}$. The tagged photons were incident on a 0.52$\pm$0.01~mm thick isotopically enriched (99.5\%) $^{208}$Pb target placed at the centre of the CB detector. The CB is a 672 element NaI detector covering ~94\% of 4$\pi$ steradians. A central detector system provided charged particle identification~\cite{pid} and track information~\cite{mwpc} and allowed the target position to be determined within $\sim$0.5~mm. 

\begin{figure}
\includegraphics[scale=0.45]{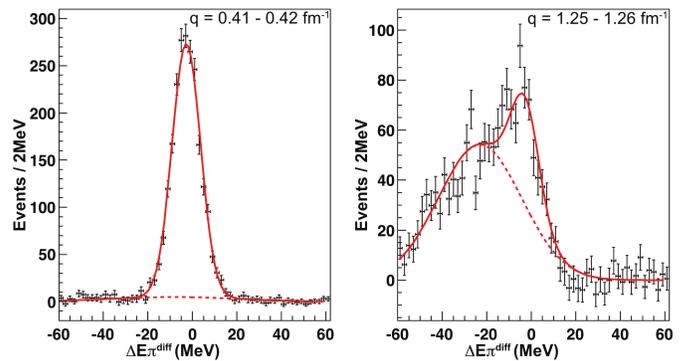}
\caption{\label{fig:imass} The fits to the spectrum of  $\Delta E_{\pi}^{diff}$ for $E_{\gamma}$=200~MeV for a momentum transfer near the 1st diffraction maximum (left) and the 1st diffraction minimum (right). 
}
\end{figure}
Neutral pions were identified in the CB from their 2$\gamma$ decay and their momenta were reconstructed from the energies and angles of the decay photons~\cite{incoh}. 
The coherent events were isolated from background by using the energy difference $\Delta E_{\pi}^{diff}$ defined as: 

\begin{equation}
\Delta E_{\pi}^{diff} = E_{\pi}^{CM} - E_{\pi}^{Det}. 
\end{equation}
$E_{\pi}^{CM}$ is the energy of the pion in the centre-of-mass (CM) frame of the incident photon and nucleus at rest, calculated using the incident photon energy assuming coherent $\pi^{0}$ production from a $^{208}$Pb nucleus. $E_{\pi}^{Det}$ is the detected $\pi^{0}$ energy in the CM frame. For a coherent reaction $\Delta E_{\pi}^{diff}$ should be close to zero. Example spectra for $\Delta E_{\pi}^{diff}$ are shown in Fig.~\ref{fig:imass}. In the first maximum of the $\pi^{0}$ angular distribution the coherent process dominates and allows determination of the width of the coherent peak. The measured $E_{\pi}^{diff}$ resolution ranged from 2~MeV near threshold to 9~MeV at $E_{\gamma}$=240~MeV, in excellent agreement with a Geant4~\cite{geant4} (G4) simulation. Near the diffraction minima a background arising from one or more non-coherent processes is evident. An additional Gaussian term in the fit gave a good description of the background which exhibited an $E_{\gamma}$ and $\theta_{\pi}$ dependence consistent with a simple Monte Carlo model of quasi-free $\pi^{0}$ production. The area of the Gaussian fitted to the coherent peak is taken as a measure of the coherent yield. 


 To obtain cross sections the yield was corrected for the $\pi^{0}$ detection efficiency. This is calculated by analysing pseudo-data from a G4 simulation of the detector apparatus using the same analysis procedure as for the real data. The detection efficiency shows no sharp dependence on pion angle and was typically around 40\%, a factor of over 30 improvement on previous measurements.  The yield was also corrected for the photon tagging efficiency ($\sim$40\%), with the procedure described in~\cite{tagg}. 
The contribution of pions not originating from the $^{208}$Pb target was found to be less than $\sim$1\% in additional runs with the target removed and was subtracted from the yield.

\begin{figure}
\includegraphics[scale=0.4]{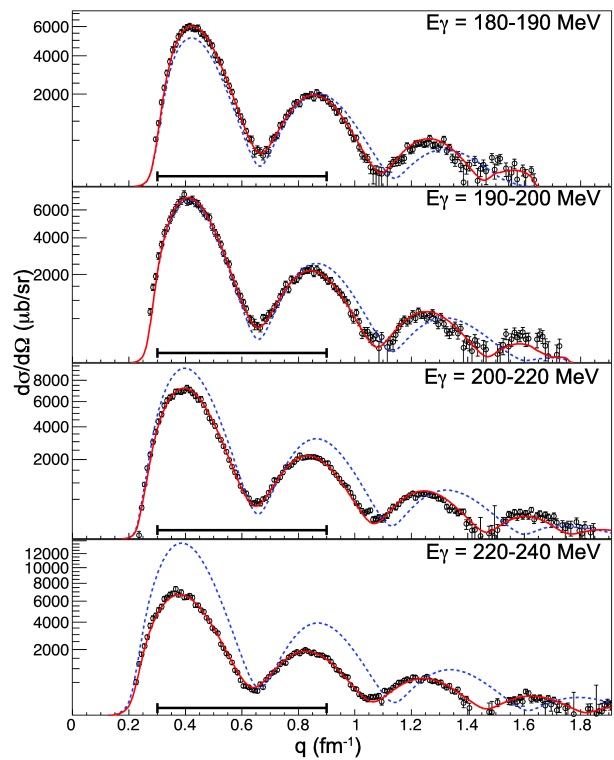} 

\caption{\label{fig:cross}Differential cross sections for the reaction $^{208}$Pb($\gamma,\pi^{0})^{208}$Pb (black circles) for the $E_{\gamma}$ regions indicated. The y-axis employs a square root scale to improve the clarity.  The red solid line shows the interpolated fit of the theoretical model to the data. The $q$ range of the fit is indicated by the horizontal bar. The dashed blue line shows the model predictions without including the pion-nucleus interaction.}
\end{figure}

The differential cross sections are analysed in terms of the momentum transfer $q$, defined as $\bf{q} = \bf{P_{\gamma}} - \bf{P_{\pi}}$ where $ {\bf P_{\gamma}}$ is the incident photon momentum and ${\bf P_{\pi}}$ is the measured pion momentum. The differential cross sections as a function of momentum transfer are presented in Fig.~\ref{fig:cross} for four $E_{\gamma}$ bins from 180 to 240~MeV. For this $E_{\gamma}$ region pion photoproduction models show agreement for the ratio of $\pi^{0}$ production from the proton and neutron to within $\pm$5\%~\cite{md07,said,BnGa}. The maximum photon energy restricts the data to regions of pion momenta where the model of Ref.~\cite{cohpi_model} predicts that FSI effects are fairly small. 
In order to extract information about the nucleon distribution in $^{208}$Pb, the measured $(\gamma,\pi^{0})$ cross sections are compared with predictions from the model of Drechsel \textit{et. al.}~\cite{cohpi_model}, which represents $\pi^{0}$ photoproduction using a unitary isobar model and includes a self-energy term for $\Delta$ propagation effects in the nucleus. The pion-nucleus interaction is treated using a complex optical potential~\cite{gmitro}, whose parameters are fixed by fits to pion-nucleus scattering data.  The model gave good agreement with coherent data from a range of nuclei~\cite{bernd2002}. In the $(\gamma,\pi^{0})$ model the nucleon density distibution, $\rho$(r), is parameterized as a single symmetrised two parameter Fermi distribution (2pF)~\cite{gmitro} with half-height radius $c$ and diffuseness $a$. For the present analysis different proton and neutron distributions, each separately parameterized by a 2pF distribution are needed to describe the nuclear shape, $\rho(r) = $(Z/A) $\rho_{p}(r)$ + (N/A) $\rho_{n}(r)$. Then in order to put $\rho(r)$ into the $(\gamma,\pi^{0})$ code it is fitted by a single 2pF distribution~\cite{ff}. The parameters for $\rho_{p}(r)$ are well determined by electron scattering~\cite{chargeradius}, viz. $a_{p}$ =  0.447 fm and $c_{p}$ = 6.680 fm. The values used have been corrected for the finite size of the proton to give the point charge distribution which is relevant for pion photoproduction~\cite{antip1}. For the neutron distribution parameters a grid of 35 points covering the ranges $c_{n}$ = 6.28 to 7.07 fm and $a_{n}$ = 0.35 to 0.65 fm was selected and the $(\gamma,\pi^{0})$ cross section was calculated at each point. These cross sections were smeared with the experimental $q$ resolution, $\sigma_{q}$ = 0.02-0.03 fm$^{-1}$ depending on E$_{\gamma}$, as determined from the G4 simulation. A two-dimensional interpolation between the 35 smeared cross sections was then used to fit the ($\gamma.\pi^{0}$) cross sections in Fig.~\ref{fig:cross} in the region $q$ = 0.3 to 0.9 fm$^{-1}$ and thus extract the best fit values $a_{n}$ and $c_{n}$ for the neutron distribution for each photon energy bin. Due to the extraordinary statistical accuracy in the maxima, an additional 3\% error was assigned to each cross section value to ensure that the important information in the first minimum was given sufficient weight in the fit. The fitted theoretical cross sections are shown together with the experimental data for all measured $q$ in Fig.~\ref{fig:cross}. Excellent fits are obtained in the fitted $q$ range with $\chi^{2}$ per degree of freedom of $\sim$1. Outside this fitted range the model still gives a very good description of the experimental data with discrepancies only evident at high $q$, probably due to the inability of the 2pF parameterisation to describe the fine details of the distribution. A normalisation parameter was included in the fit and was found to vary only within $\pm$5\% of unity. Figure 2 also shows model predictions when the $\pi^{0}$-nucleus interaction is not included in the calculation. For the three lower $E_{\gamma}$ bins the effect of the $\pi^{0}$-nucleus interaction is modest in the fitted $q$ range. For the highest bin the differences are significantly larger, probably due to the increase in the $\pi^{0}$ absorption cross section for $\pi^{0}$ energies in the region of the $\Delta$ resonance.  
\begin{figure}
\includegraphics[scale=0.45]{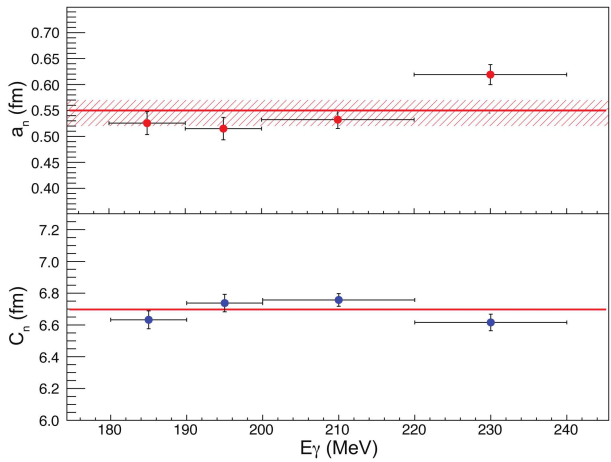} 

\caption{\label{fig:fit}The best fit for diffuseness (upper) and half-height radius (lower) for the neutron distribution for each $E_{\gamma}$ bin.  The horizontal lines show the average over $E_{\gamma}$ and the shaded band shows the estimated systematic error (see text)}
\end{figure}

The best fit half-height radius and diffuseness parameters for the neutron distribution are plotted for each $E_{\gamma}$ bin in Fig. \ref{fig:fit}. The solid lines show the average values. The half-height radii are statistically consistent with the average. For the diffuseness the value obtained from the highest $E_{\gamma}$ bin has a 3.5$\sigma$ variation. From Fig. 2 it is clear that for this $E_{\gamma}$ bin the effects of the $\pi^{0}$-nucleus interaction are predicted to be largest. This may lead to larger systematic errors, particularly for the diffuseness that is constrained by the relative heights of the 1st and 2nd maxima. An estimate of the systematic error was obtained from the variation in the average with and without the inclusion of this point. This variation ($\sim$0.025~fm) is shown by the shaded band in Fig.~\ref{fig:fit}. Additional estimates of the systematic uncertainty in the half-radii were obtained from analysis using only the first minimum and varying the diffuseness over the range $a_{n}$=0.53-0.59 predicted by models~\cite{centelles}. This method produced consistent results. Variation of the relative weighting of the proton and neutron amplitudes by 10\%, consistent with the estimated uncertainty in the pion production amplitudes, gave a change of $\sim$0.02~fm in the diffuseness and had a negligible effect on the half-height radius. Variation in the modelling of the background used to extract the coherent strength gave a systematic error in the diffuseness of $\sim$0.01~fm. The half-height radius and diffuseness of the neutron distribution are found to be 6.70$\pm0.03(stat.)$~fm and 0.55$\pm0.01(stat.)$$^{+0.02}_{-0.03}(sys.)$~fm respectively. This corresponds to a neutron skin thickness $\Delta r_{np}$=0.15$\pm0.03(stat.)$$^{+0.01}_{-0.03}(sys.)$~fm, obtained using the analytic relationship between  $\Delta r_{np}$ and the 2pF parameters~\cite{2pfskin}. Although slightly smaller than the result~\cite{ion_diff} from heavy ion work the present result is in
good  agreement with the other  previous measurements~\cite{Tsang2,zenihiro,friedman,antip1,antip2,Carbone,kras,tamii,Reinhard}.

\begin{figure}
\includegraphics[scale=0.45]{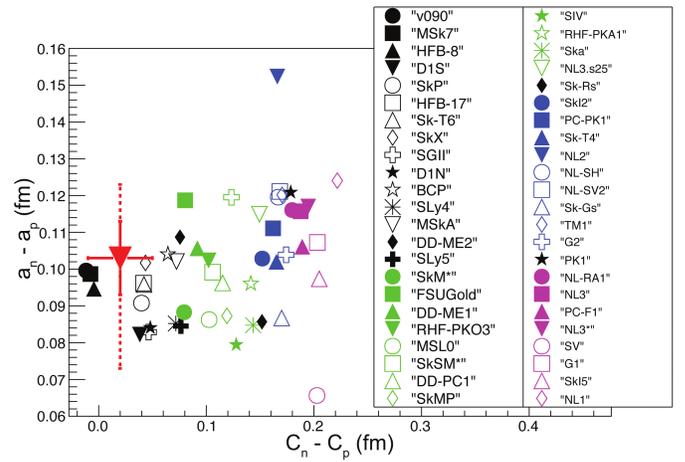}

\caption{\label{fig:skin} The quantity ($a_{n}$-$a_{p}$) plotted versus ($c_{n}$ -$c_{p}$) for $^{208}$Pb. The red inverted triangle shows the present result with statistical and systematic errors shown by solid and dashed lines respectively. The symbols show the predictions from the various nuclear structure calculations, obtained from Ref.~\cite{centelles_2}.}
\end{figure}
The new results are compared to the current predictions from nuclear structure models in Fig \ref{fig:skin}, adopting the framework from Ref.~\cite{centelles_2}, where the theoretical model predictions are fitted with a 2pF function. The difference between the diffuseness parameters for neutrons and protons is plotted versus the difference in half-height radii. The present result clearly shows that the diffuseness of the neutron distribution in $^{208}$Pb is in the range of theoretical predictions and is significantly larger than that for the protons. This is an important result as conclusive experimental evidence for a difference in the diffuseness of the nuclear density distributions has been lacking~\cite{centelles_droplet}. A pure ``skin'' effect would have $a_{n}$-$a_{p}$ close to zero, so this new work indicates the neutron ``skin'' of $^{208}$Pb has a halo character. 

In summary, a measurement of the coherent photoproduction of $\pi^{0}$ mesons from $^{208}$Pb has provided the first determination of a nuclear matter form factor with an electromagnetic probe. The existence of a neutron skin on the surface of the $^{208}$Pb nucleus is confirmed with a thickness $\Delta r_{np}$=0.15$\pm 0.03(stat.)$$^{+0.01}_{-0.03}(sys.)$~fm. The method is sensitive enough to extract the shape of the neutron distribution, which is found to be $\sim$20\% more diffuse than the charge distribution. This new determination of the neutron skin properties discriminates against some of the modern nuclear theories in common use and will be a valuable new constraint on the equation of state for neutron-rich matter and neutron stars. 
\begin{acknowledgments}
The authors wish to acknowledge the excellent support of
the accelerator group of MAMI. This work was supported
by the UK STFC the Deutsche Forschungsgemeinschaft (SFB
443), SFB/Transregio16, Schweizerischer Nationalfonds and the European Community-Research Infrastructure Activity FP6 and FP7, the USDOE, USNSF and NSERC (Canada). 
\end{acknowledgments}






\begin{thebibliography}{99}
\bibitem{ff}To check the validity of approximating $\rho(r)$ by a single 2pF distribution, the Fourier transforms of $\rho(r)$ and the fitted 2pF distribution were compared. In the momentum range up to $q$=0.9~fm$^{-1}$, over which the ($\gamma,\pi^{0}$) cross section was fitted to extract information about $\rho_{n}(r)$, the fractional difference between the two transforms only rises to 0.3\% at $q$=0.9~fm$^{-1}$.
\end{thebibliography}
\end{document}